\documentclass[a4paper,pra,twocolumn,amssymb,amsfonts,floatfix,superscriptaddress,longbibliography,notitlepage,footinbib,accepted=2022-01-27]{quantumarticle}
\pdfoutput=1
\usepackage[utf8]{inputenc}
\usepackage{color}
\usepackage{float}
\usepackage{xcolor}
\usepackage{mathtools,amsmath,amsthm,amsfonts,amssymb}
\usepackage{commath}
\usepackage{physics}
\usepackage{bm}
\usepackage[pdftex]{graphicx}
\usepackage{hyperref}
\usepackage{enumitem}
 
\begin{document}

\DeclarePairedDelimiterX{\energyshellaverage}[2]{\langle}{\rangle_{\displaystyle #2}}{#1}
\DeclarePairedDelimiterX{\energyshellaveragelong}[3]{\langle}{\rangle_{#2\in\mathcal{M}(#3)}}{#1}

\title{Identification of quantum scars via phase-space localization measures}

\author{Sa\'ul Pilatowsky-Cameo}
\affiliation{Instituto de Ciencias Nucleares, Universidad Nacional Aut\'onoma de M\'exico, Apdo. Postal 70-543, C.P. 04510 CDMX, Mexico}
\affiliation{Center for Theoretical Physics, Massachusetts Institute of Technology, Cambridge, MA 02139, USA
}
\author{David Villase\~nor}
\affiliation{Instituto de Ciencias Nucleares, Universidad Nacional Aut\'onoma de M\'exico, Apdo. Postal 70-543, C.P. 04510 CDMX, Mexico}
\author{Miguel A. Bastarrachea-Magnani}
\affiliation{Departamento de F\'isica, Universidad Aut\'onoma Metropolitana-Iztapalapa, San Rafael Atlixco 186, C.P. 09340 CDMX, Mexico}
\author{Sergio Lerma-Hern\'andez}
\affiliation{Facultad de F\'isica, Universidad Veracruzana, Circuito Aguirre Beltr\'an s/n,  C.P. 91000 Xalapa, Veracruz, Mexico}
\author{Lea~F.~Santos} 
\affiliation{Department of Physics, Yeshiva University, New York, New York 10016, USA}
\author{Jorge G. Hirsch}
\affiliation{Instituto de Ciencias Nucleares, Universidad Nacional Aut\'onoma de M\'exico, Apdo. Postal 70-543, C.P. 04510  CDMX, Mexico}


\begin{abstract}
There is no unique way to quantify the degree of delocalization of quantum states in unbounded continuous spaces. In this work, we explore a recently introduced localization measure that quantifies the portion of the classical phase space occupied by a quantum state. The measure is based on the $\alpha$-moments of the Husimi function and is known as the R\'enyi occupation of order $\alpha$. With this quantity and random pure states, we find a general expression to identify states that are maximally delocalized in phase space. Using this expression and the Dicke model, which is an interacting spin-boson model with an unbounded four-dimensional phase space, we show that the R\'enyi occupations with $\alpha>1$ are highly effective at revealing quantum scars. Furthermore, by analyzing the high moments ($\alpha>1$) of the Husimi function, we are able to identify qualitatively and quantitatively the unstable periodic orbits that scar some of the eigenstates of the model.
\end{abstract} 

\maketitle

\section{Introduction}

In the classical domain, typical trajectories of chaotic systems fill the available phase space.  Even if unstable periodic trajectories are present, one does not find them by random sampling the phase space, because they form a measure-zero set. In the quantum domain, however, their existence  is  revealed through quantum scars. These are  states that  are  not uniformly distributed, but present high probabilities along  the  phase-space  region  occupied  by  unstable periodic  trajectories.  The  unexpected  phenomenon  of  quantum  scarring was observed in early studies of the Bunimovich stadium billiard~\cite{McDonald1979,McDonald1983}, although the term was coined later~\cite{Heller1984,Heller1987,Berry1989,Heller1991,Kus1991}. Recently, the subject has received boosted attention due to the so-called ``many-body quantum scars'' ~\cite{Turner2018,Turner2021}, which still await for possible links with the classical phase space.

Quantum scarring and localization are not synonyms~\cite{Pilatowsky2021NatCommun}, but 
both carry the idea of confined eigenstates. The notion of localization in quantum systems presupposes a basis representation. Anderson localization~\cite{Anderson1958,Lee1985}, for example, refers to the suppression of the classical diffusion of particles in real space due to quantum interferences. This phenomenon has a dynamical counterpart originally studied in the context of the kicked rotor~\cite{Chirikov1981,Fishman1982,Izrailev1990,Fishman2010}. Localization is an extremely broad subject that extends to the hydrogen atom in a monochromatic field~\cite{Casati1984a,Casati1987}, Rydberg atoms~\cite{Blumel1987}, quantum billiards~\cite{Borgonovi1996,Batistic2010,Batistic2013a,Batistic2013b,Porter2017,Batistic2019,Robnik2020}, banded random matrices~\cite{Casati1993}, and interacting many-body quantum systems~\cite{SantosEscobar2004,Basko2006,Oganesyan2007,Lazarides2015,Stanley2021}, among others.  

In finite spaces, the degree of delocalization of a quantum state is based on how much it spreads on a chosen finite basis representation, as quantified by participation ratios~\cite{Edwards1972,Izrailev1990} and R\'enyi entropies~\cite{Atas2012,Atas2014}. In unbounded continuous spaces, such measures can reach arbitrarily large values. If one wishes to investigate maximally delocalized states in these systems, one must define a finite region as a reference volume. There are multiple ways to select this bounded region, and each choice produces a different localization measure~\cite{Villasenor2021, Wang2020}. A natural volume of reference is obtained by measuring the degree of localization of a quantum state over the classical phase-space energy shell, as proposed in Ref.~\cite{Villasenor2021}. This measure of localization has been named R\'enyi  occupation, because it is related to the exponential of the R\'enyi-Wehrl entropies~\cite{Wehrl1978,Gnutzmann2001}. To compute the R\'enyi occupation, we represent the quantum states in the phase space through   the Husimi function~\cite{Husimi1940}. Thus, the R\'enyi  occupations of order $\alpha$ refer to averages of normalized $\alpha$-moments of the Husimi functions.

In this work, we obtain a general analytical expression for the R\'enyi occupations of order $\alpha$ using maximally delocalized random pure states. We employ this expression to analyze the eigenstates of the Dicke model in the chaotic regime and distinguish the eigenstates that are maximally delocalized from those that are highly localized in phase space. The Dicke model~\cite{Dicke1954} is an experimental spin-boson model that has a rich and unbounded phase space.

By comparing the R\'enyi occupations of the high-energy eigenstates of the Dicke model with the R\'enyi occupations of random states, as a function of $\alpha$, we range the eigenstates according to their degree of delocalization. While all eigenstates are scarred~\cite{Pilatowsky2021NatCommun}, the most localized states are the ones scarred by unstable periodic orbits with short periods. The analysis of high-moments ($\alpha>1$) of the Husimi functions of these highly localized states allow us to identify more clearly the classical unstable periodic orbits that cause their scarring. These orbits are different from the ones found in Ref.~\cite{Pilatowsky2021} and their uncovering represents a step forward in the difficult task of identifying the unstable periodic orbits underlying  quantum scars. In addition, we study the dynamical properties of the unveiled orbits and use them to explain the structures of the eigenstates and their degree of localization.

The paper is organized as follows. The Dicke model and its phase space are presented in Sec.~\ref{sec:DickeModel}.  In Sec.~\ref{sec:RenyiOccupation}, we introduce a measure of localization of quantum states based on the R\'enyi occupations of order $\alpha$ and provide an analytical expression for maximally delocalized states. In Sec.~\ref{sec:3} we discuss quantum scarring in the light of the R\'enyi occupations and identify the unstable  periodic orbits that scar the most localized eigenstates. We also interpret our localization measure  in terms of the geometric and dynamical properties of the classical periodic orbits. Our conclusions are presented in Sec.~\ref{sec:Conclusions}.

\section{Dicke Model}
\label{sec:DickeModel}

The Dicke model~\cite{Dicke1954}  was initially introduced to explain the phenomenon of superradiance~\cite{Hepp1973a,Wang1973,Garraway2011,Kirton2019} and has since fostered a wide variety of theoretical studies, including the behavior of out-of-time-ordered correlators~\cite{Chavez2019,Lewis-Swan2019,Pilatowsky2020}, manifestations of quantum scarring~\cite{Deaguiar1992,Furuya1992,Bakemeier2013,Pilatowsky2021NatCommun,Pilatowsky2021}, non-equilibrium dynamics~\cite{Altland2012NJP,Kloc2018,Lerma2018,Lerma2019,Kirton2019,Villasenor2020}, and  measures of quantum localization with respect to phase space~\cite{Wang2020,Pilatowsky2021NatCommun}. The model is also of great interest to experiments with trapped ions~\cite{Cohn2018,Safavi2018}, superconducting circuits~\cite{Jaako2016}, and cavity assisted Raman transitions~\cite{Baden2014,Zhang2018}.

The Dicke Hamiltonian  describes a single mode of the electromagnetic field interacting with a set of $\mathcal{N}$ two-level atoms~\cite{Dicke1954}. Setting $\hbar=1$, the Hamiltonian is given by
\begin{equation}
\label{eqn:qua_hamiltonian}
\hat{H}_{D}=\omega\hat{a}^{\dagger}\hat{a}+\omega_{0}\hat{J}_{z}+\frac{\gamma}{\sqrt{\mathcal{N}}}(\hat{a}^{\dagger}+\hat{a})(\hat{J}_{+}+\hat{J}_{-}),
\end{equation}
where $\hat{a}^{\dagger}$ ($\hat{a}$) is the bosonic creation (annihilation) operator of the field mode and $\hat{J}_{+}$ ($\hat{J}_{-}$) is the raising (lowering) operator defined by $\hat{J}_{\pm}=\hat{J}_{x}\pm i\hat{J}_{y}$, where $\hat{J}_{x,y,z}=(1/2)\sum_{k=1}^{\mathcal{N}}\hat{\sigma}_{x,y,z}^{k}$ are the collective pseudo-spin operators satisfying the SU(2) algebra and $\hat{\sigma}_{x,y,z}$ are the Pauli matrices. The Hamiltonian parameters are the radiation frequency of the electromagnetic field $\omega$, the transition frequency of the two-level atoms $\omega_{0}$, and the atom-field coupling strength $\gamma$. Since the Hamiltonian commutes with the total pseudo-spin operator $\hat{\textbf{J}}^{2}=\hat{J}_{x}^{2}+\hat{J}_{y}^{2}+\hat{J}_{z}^{2}$, the Hilbert space is separated into different invariant subspaces for each eigenvalue $j(j+1)$ of $\hat{\textbf{J}}^{2}$. We work within the totally symmetric subspace that includes the ground-state of the system and is defined by the maximum value of the pseudo-spin length $j=\mathcal{N}/2$. The model also possesses  a discrete parity symmetry, $\hat{\Pi}=e^{i \pi (\hat{a}^{\dagger} \hat{a}+ \hat{J}_z+j)}$, which leads to an additional subspace separation according to the eigenvalues of the parity operator $\Pi_\pm=\pm 1$.   

The model has been used in studies of the quantum phase transition from a normal ($\gamma<\gamma_c$) to a superradiant ($\gamma>\gamma_c$) phase, which arises when the coupling strength reaches the critical value $\gamma_{c}=\sqrt{\omega\omega_{0}}/2$~\cite{Hepp1973a,Hepp1973b,Wang1973,Emary2003}. The model also exhibits regular or chaotic behavior depending on the Hamiltonian parameters and excitation energies~\cite{Chavez2016}. In this paper, we choose the resonant frequency case $\omega=\omega_0=1$, system size $j=\mathcal{N}/2=100$, and $\gamma=2\gamma_c$ to work in the superradiant phase. We  focus on high energies, where the system displays hard-chaos behavior. 

\subsection{Classical Limit}

The classical Hamiltonian of the Dicke model is obtained by taking the expectation value of the quantum Hamiltonian $\hat{H}_{D}$ under the tensor product of bosonic Glauber and atomic Bloch coherent states, that is $|\bm x\rangle=|q,p\rangle\otimes~|Q,P\rangle$~\cite{Deaguiar1991,Deaguiar1992,Bastarrachea2014a,Bastarrachea2014b,Bastarrachea2015,Chavez2016,Villasenor2020}, and dividing it by the system size,
\begin{align}
h_\text{cl}(\bm x)  &=\frac{\langle\bm{x}|\hat{H}_{D}|\bm{x}\rangle}{j} =\frac{\omega}{2}\big(p^{2}+q^{2}\big)+\\&\nonumber+\frac{\omega_{0}}{2}\big(P^2+Q^2\big)+2\gamma q Q\sqrt{1-\frac{P^2+Q^2}{4}}-\omega_{0}.
\end{align} 
The bosonic Glauber and  atomic Bloch coherent states are, respectively, 
\begin{align}
|q,p\rangle & =e^{-(j/4)\left(q^{2}+p^{2}\right)}\exp(\sqrt{\frac{j}{2}}\left(q+ip\right)\hat{a}^{\dagger})|0\rangle, \\
|Q,P\rangle & =\bigg(1-\frac{P^2+Q^2}{4}\bigg)^{j}\!\!\exp(\frac{\left(Q+iP\right)\hat{J}_{+}}{\scriptstyle\sqrt{4-P^2-Q^2}})|j,-j\rangle, \nonumber
\end{align}
with $|0\rangle$ being the photon vacuum and $|j,-j\rangle$ the state with all the atoms in their ground state. 

\subsection{Phase space}

The classical Hamiltonian $h_\text{cl}(\bm x)$ has a four-dimensional phase space $\mathcal{M}$ in the coordinates $\bm x=(q,p;Q,P)$, where $P^2 + Q^2 \leq 4$, because the pseudo-spin degree of freedom is bounded. This space can be partitioned into a family of classical energy shells
\begin{equation*}
    \mathcal{M}(\epsilon)=\{\bm x\in \mathcal{M} \,\mid\, h_\text{cl}(\bm x)=\epsilon\} 
\end{equation*}
in terms of the rescaled classical energy $\epsilon=E/j$. Due to energy conservation, for a given initial condition $\bm x\in \mathcal{M}(\epsilon)$, the classical evolution given by $h_\text{cl}$ remains in $\mathcal{M}(\epsilon)$, that is, $\bm x(t)\in  \mathcal{M}(\epsilon)$ for all times $t$. 

The classical energy shells are bounded with respect to the three-dimensional surface measure $\dif \bm x\, \delta(h_\text{cl}(\bm x) - \epsilon)$. The finite volume of $\mathcal{M}(\epsilon)$ is given by 
\begin{equation}
    \int_{\mathcal{M}}\dif \bm x\,  \delta(h_\text{cl}(\bm x)-\epsilon) =(2\pi\hbar_{\text{eff}})^2\nu(\epsilon),
\end{equation}
 where  $\hbar_{\text{eff}}=1/j$ is the effective Planck constant~\cite{Ribeiro2006}, which determines the volume of the Plank cell $(2\pi \hbar_\text{eff})^2$, and $\nu(\epsilon)$ is the semiclassical density of states obtained by taking only the first term in the Gutzwiller trace formula \cite{Gutzwiller1971,Gutzwiller1990book,Bastarrachea2014a} (see App.~\ref{app:SDoS} for more details on this quantity).
 
To calculate the average value, $\energyshellaverage*{f}{\epsilon}$,  of a  phase-space function $f(\bm x)$ over the classical energy shell $\mathcal{M}(\epsilon)$, we integrate with respect to the surface measure $\dif \bm x\, \delta(h_\text{cl}(\bm x) - \epsilon)$ and divide the result by the total volume $(2\pi\hbar_{\text{eff}})^2\nu(\epsilon)$,
\begin{align}
\label{av:definition}
    \energyshellaverage*{f}{\epsilon}&=\energyshellaveragelong*{f(\bm x)}{\bm x}{\epsilon}\\&\equiv \frac{1}{(2\pi\hbar_{\text{eff}})^2\nu(\epsilon)}\int_{\mathcal{M}}\!\!\dif \bm x\,  \delta(h_\text{cl}(\bm x)-\epsilon) f(\bm x). \nonumber
\end{align}

\section{R\'enyi Occupations}
\label{sec:RenyiOccupation}

The R\'enyi entropy of order $\alpha \geq 0$ \cite{Renyi1961},
\begin{equation}
\label{eq:renyi}
S_{\alpha} (\mathcal{B},\hat{\rho})= \frac{1}{1-\alpha} \log \left(\sum_{i=1}^\infty \expval{\hat{\rho}}{\phi_i}^\alpha\right),
\end{equation}
is a common tool to measure the degree of delocalization of a quantum state $\hat{\rho}$ over a given basis $\mathcal{B}=\{\ket{\phi_i}\mid i\in \mathbb{N}\}$ that forms a countable set of states. The quantity $L_\alpha = \exp(S_\alpha)$ is the generalized participation ratio to the power $1/(\alpha -1)$, and it counts the effective number of states $\ket{\phi_i}$ that compose the state $\hat{\rho}$. The fact that $\mathcal{B}$ is an orthonormal basis ensures that $ \sum_{i=1}^\infty \expval{\hat{\rho}}{\phi_i}=1$. If $\mathcal{B}$ is an arbitrary countable set of states, the measure $L_\alpha$ can be generalized by performing an additional normalization as follows:
\begin{equation}
\label{eq:PRGen}
   L_\alpha(\mathcal{B},\hat{\rho})=\left(\frac{\sum_{i=1}^\infty \expval{\hat{\rho}}{\phi_i}^\alpha}{\left(\sum_{i=1}^\infty  \expval{\hat{\rho}}{\phi_i}\right)^\alpha}\right)^{1/(1-\alpha)}. 
\end{equation}

The same idea can be extended to the case of non-countable sets of states. Consider the set 
\begin{equation}
\mathcal{C}_\epsilon=\{\ket{\bm x}\mid \bm x\in \mathcal{M}(\epsilon)\}
\end{equation}
 of coherent states over a classical energy shell $\mathcal{M}(\epsilon)$ at energy $\epsilon$. In this case,  a measure similar to that in Eq.~(\ref{eq:PRGen}) can be defined, written in terms of the Husimi function $\mathcal{Q}_{\hat{\rho}}$. This function is a quasi-distribution used to represent a quantum state in the phase space $\mathcal{M}$, and is given by
\begin{equation}
    \mathcal{Q}_{\hat{\rho}}(\bm x)=\expval{\hat{\rho}}{\bm x}\geq 0
\end{equation}
for each point $\bm x\in{\mathcal{M}}$. 

For a non-countable set of states we can no longer perform a discrete sum over the states of $\mathcal{C}_\epsilon$, but we can instead use the three-dimensional surface measure of $\mathcal{M}(\epsilon)$ to  perform an average over $\mathcal{M}(\epsilon)$, as defined in Eq.~\eqref{av:definition},
 \begin{equation}
 \label{eq:HuAverages}
     \energyshellaverage{\mathcal{Q}_{\hat{\rho}}^\alpha}{\epsilon}=\energyshellaveragelong{\expval{\hat{\rho}}{\bm x}^\alpha}{\bm x}{\epsilon}.
 \end{equation} 
 Substituting the sums $\sum_{i=1}^\infty \expval{\hat{\rho}}{\phi_i}^\alpha$ in Eq.~\eqref{eq:PRGen} by these averages, we obtain the R\'enyi occupations  of order $\alpha \geq 0$~\cite{Villasenor2021},
\begin{equation}
\label{eq:Lalphadefinition}
    \mathfrak{L}_\alpha(\epsilon,\hat{\rho})=\left(\frac{\energyshellaverage[\big]{\mathcal{Q}_{\hat{\rho}}^{\alpha}}{\epsilon}}{\energyshellaverage*{\mathcal{Q}_{\hat{\rho}}}{\epsilon}^{\alpha}}\right)^{1/(1-\alpha)}.
\end{equation}
 In the limit  $\alpha\to 1$, Eq.~\eqref{eq:Lalphadefinition} gives
\begin{equation}
    \mathfrak{L}_1(\epsilon,\hat{\rho})=\energyshellaverage*{\mathcal{Q}_{\hat{\rho}}}{\epsilon}\exp(-\frac{\energyshellaverage*{\mathcal{Q}_{\hat{\rho}}\log \mathcal{Q}_{\hat{\rho}}}{\epsilon}}{\energyshellaverage*{\mathcal{Q}_{\hat{\rho}}}{\epsilon}}).
\end{equation}
The name ``R\'enyi occupations'' is inspired by the relationship of these measures with the exponential of the R\'enyi entropies. The values of $\mathfrak{L}_\alpha(\epsilon,\hat{\rho})$ range from 0 to 1, and they indicate the percentage of the classical energy shell that is occupied by the quantum state. The R\'enyi occupation of a quantum state equals 1 if the corresponding Husimi function is constant in the classical energy shell, which means that the state is uniformly delocalized, covering the classical energy shell homogeneously. Notice, however, that this value is not reached by any realistic pure state and not even by random pure states, as explained in the subsection below.

In this work, we explore how the R\'enyi occupation depends on $\alpha$ and which information about the degree of localization of the quantum state can be  extracted from different $\alpha$'s. In general, $\alpha<1$  makes the Husimi distribution look more homogeneous over the classical energy shell. In the limit $\alpha=0$, one has
\begin{equation}
 \mathfrak{L}_0(\epsilon,\hat{\rho})=\frac{\energyshellaverage*{\mathcal{Q}_{\hat{\rho}}^{0}}{\epsilon}}{\energyshellaverage*{\mathcal{Q}_{\hat{\rho}}}{\epsilon}^{0}}=\frac{\energyshellaverage*{1}{\epsilon}}{1}=1
\end{equation}
for any state, because the Husimi function can be zero only on a zero-measure set of points. In contrast, for $\alpha>1$, the R\'enyi occupation becomes less sensitive to the regions of the classical energy shell where the Husimi function is relatively small. As $\alpha$ increases, these regions cease to contribute to $ \mathfrak{L}_{\alpha}(\epsilon,\hat{\rho})$, leaving only those regions where the Husimi function attains relatively large values, and thus decreasing the occupation value. We return to this discussion at the end of Sec.~\ref{secIIIC} and on how to make use of the $\alpha$-moments of the Husimi function in the analysis of scarring in Sec.~\ref{sec:3}. 

To benchmark the R\'enyi occupations as a measure of localization, we describe below the behavior of $\mathfrak{L}_\alpha(\epsilon,\hat{\rho}_R)$ for random pure states $\hat{\rho}_R = \ket{\psi_R} \bra{\psi_R}$, which are the most delocalized states over the phase space. For this, we need to investigate the Husimi moments  $\energyshellaverage{\mathcal{Q}_{\hat{\rho}_R}^{\alpha}}{\epsilon}$, which we do first for systems with a finite Hilbert space, before proceeding with the Dicke model.

\subsection{Maximally delocalized states in finite Hilbert spaces}

It has been shown that for random pure states $\ket{\psi_R}$ completely spread in a Hilbert space of dimension $N$, the statistical averages $\expval{\, \cdot \,}_{\psi_R}$ of the projections of $\ket{\psi_R}$ into an arbitrary state $\ket{\phi}$ gives~\cite{Kus1988,Jones1990,Gnutzmann2001} 
\begin{equation}
\label{eq:formulascalingOfaverage}
    \expval{\abs{\braket{\psi_R}{\phi}}^{2\alpha}}_{ \psi_R}=\frac{\Gamma(N)\Gamma(1+\alpha)}{\Gamma(N+\alpha)}
\end{equation}
where $\Gamma$ is the gamma function. Applying this result to coherent states $\ket{\phi}=\ket{\bm x}$ and performing an additional average over all points $\bm x$ of the phase space $\mathcal{M}$, we obtain
\begin{equation}
    \expval{ \expval{\abs{\braket{\psi_R}{\bm x}}^{2\alpha}}_{\bm x\in {\mathcal{M}}}}_{ \psi_R}=\frac{\Gamma(N)\Gamma(1+\alpha)}{\Gamma(N+\alpha)}.
\end{equation}
As $N$ increases, one expects that the variance of the averages $\expval{\, \cdot \,}_{\psi_R}$ will decrease \cite{Pilatowsky2021NatCommun}. Thus, for sufficiently large $N$, a single, but typical, random state $\ket{\psi_R}$ will satisfy
\begin{equation}
    \expval{\abs{\braket{\psi_R}{\bm x}}^{2\alpha}}_{\bm x\in {\mathcal{M}}}=\frac{\Gamma(N)\Gamma(1+\alpha)}{\Gamma(N+\alpha)}.
\end{equation}
Moreover, in the limit of large $N$, we can do the approximation
$ \Gamma(N+\alpha)/\Gamma(N) \approx {N^\alpha}$, which leads to the result
\begin{equation}
\label{eq:random_averages_independent_of_N}
\left(\frac{
    \Big \langle
        \abs{\braket{\psi_R}{\bm x}}^{2\alpha}
    \Big \rangle_{\bm x\in{\mathcal{M}}}
} 
{   \Big \langle
        \abs{\braket{\psi_R}{\bm x}}^{2}
    \Big \rangle_{\bm x\in{\mathcal{M}}}^\alpha
}
\right)^{1/(1-\alpha)}    \hspace{-0.6 cm } \approx  \;\;\;  \Gamma(1+\alpha)^{1/(1-\alpha)},
\end{equation}
that is independent of the dimension $N$.

\subsection{Maximally delocalized states in the Dicke model}

In analogy to Eq. \eqref{eq:random_averages_independent_of_N}, we say that an arbitrary state $\hat{\rho}=\dyad{\psi}$ of the Dicke model is maximally delocalized if
\begin{equation}
\label{eq:quotient_averages_fin_dim}
 \mathfrak{L}_\alpha(\epsilon,\hat{\rho})  \approx    \mathfrak{L}_\alpha^\text{max} \equiv\Gamma(1+\alpha)^{1/(1-\alpha)}. 
\end{equation}
This maximum level of delocalization is attained, on average, by random pure states, although several eigenstates of the Dicke model are maximally delocalized as well.

The reason why we can extrapolate the result in Eq.~\eqref{eq:random_averages_independent_of_N}, valid for large finite-dimensional Hilbert spaces, to the infinite-dimensional Hilbert space of the Dicke model is because we perform averages over individual classical energy shells, which are bounded. Their finite volume in phase space induces a large but finite effective dimension that allows us to treat the Hilbert space of the Dicke model as if it were finite dimensional~\cite{Pilatowsky2022Arxiv}. 

In our studies of the Dicke model in Ref.~\cite{Pilatowsky2021NatCommun}, we found numerically that $\mathfrak{L}_{2}^{\text{max}}=1/2$ for pure random states.  In Ref.~\cite{Batistic2020}, an approximate maximum value of $0.7$ was found numerically in billiards for a measure similar to $\mathfrak{L}_1$. These values are obtained directly from Eq.~\eqref{eq:quotient_averages_fin_dim}, which gives $\mathfrak{L}_2^\text{max}= \Gamma(3)^{-1}=1/2$ and  $\mathfrak{L}_1^\text{max}=\lim_{\alpha\to 1} \Gamma(1+\alpha)^{1/(1-\alpha)} \approx 0.66$. These results reinforce the validity of the expression~\eqref{eq:quotient_averages_fin_dim} and suggest that it may be applicable to other models.

Notice that the upper bound for the degree of delocalization of pure states given by $\mathfrak{L}_\alpha^\text{max}$ is not equal to one, instead $\mathfrak{L}_{\alpha}^{\text{max}}<1$ for $\alpha >0$. This happens because quantum interference effects prevent a quantum pure state from homogeneously covering the classical energy shell and cause its Husimi function to be zero in some points of the phase space~\cite{Korsch1997}. Only mixed states, which can be obtained by performing infinite-time averages, can homogeneously cover the classical energy shells~\cite{Pilatowsky2021NatCommun}.

\subsection{Eigenstates}
\label{secIIIC}

\begin{figure}[ht]
\centering
\includegraphics[width=1\columnwidth]{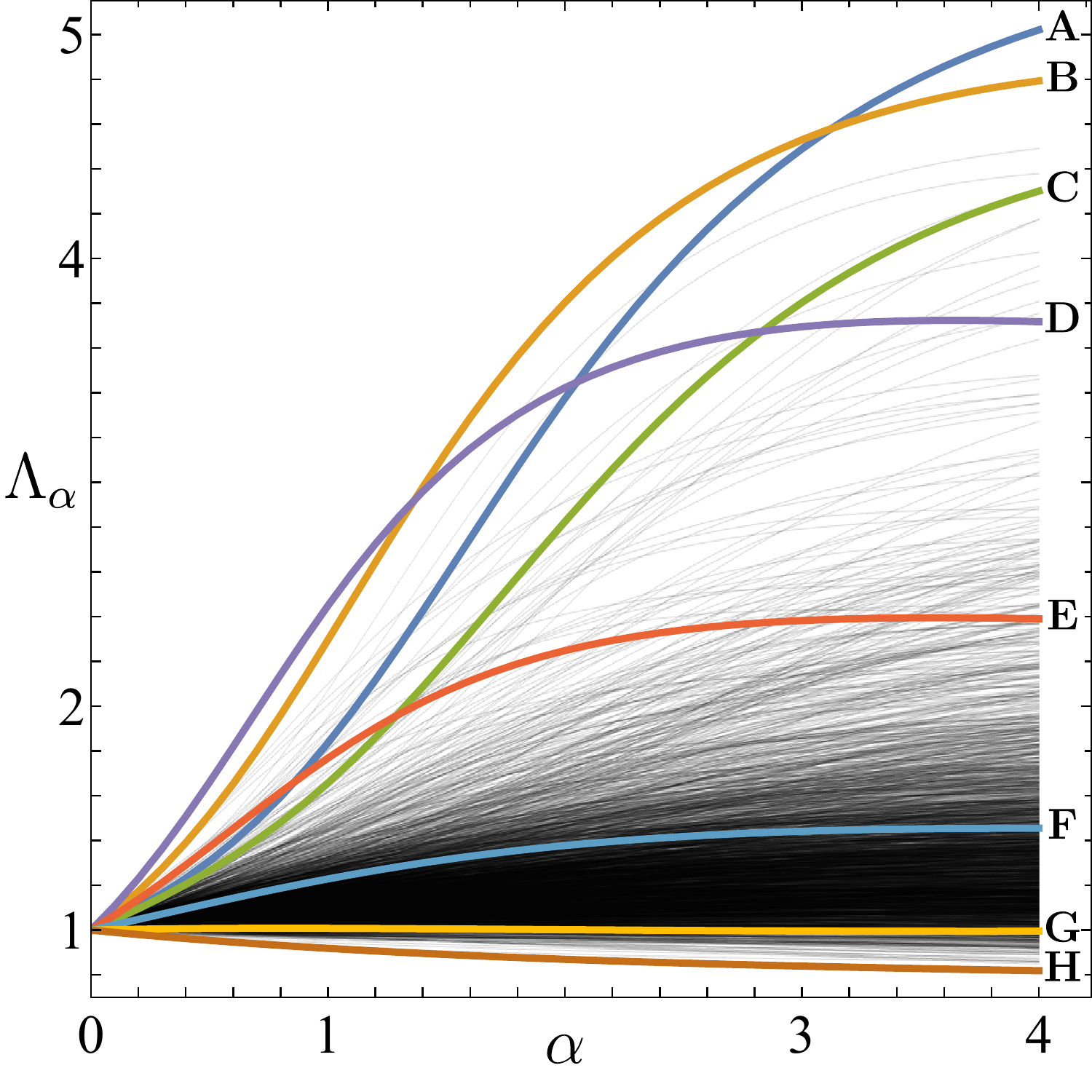}
\caption{Localization measure $\Lambda_\alpha(\epsilon_k,\hat{\rho}_k)$ for the $2437$ eigenstates $\hat{\rho}_k=\dyad{E_k}$ with eigenenergies $\epsilon_k$  inside the chaotic energy interval $[-0.6,-0.4]$. The 8 representative eigenstates labeled A-H are further analyzed in Fig.~\ref{fig02}. The system size is $j=100$.}
\label{fig01}
\end{figure}

To quantify how close the degree of delocalization of a quantum state is to the maximal value attained by random states, we define the following measure:
\begin{equation}
\label{eq:lambda}
    \Lambda_\alpha(\epsilon,\hat{\rho})=\frac{\mathfrak{L}_\alpha^\text{max}}{\mathfrak{L}_\alpha(\epsilon,\hat{\rho})}.
\end{equation}
A state that is as delocalized as a random pure state gives $\Lambda_\alpha\approx 1$, which is the lower bound for this measure. As the state $\hat{\rho}$ becomes more localized in the classical energy shell at $\epsilon$, $\Lambda_\alpha(\epsilon,\hat{\rho})$ increases. For a given $\alpha$, the value $\Lambda_\alpha(\epsilon,\hat{\rho})=n$ indicates that the state $\hat{\rho}$ is $n$ times more localized than a random state. 

In Fig.~\ref{fig01}, we study $\Lambda_\alpha (\epsilon_k,\hat{\rho}_k)$ as a function of $\alpha$ for all the eigenstates $\hat{\rho}_k=\dyad{E_k}$ of the Dicke model with eigenenergies $\epsilon_k=E_k/j$  inside the energy interval $[-0.6,-0.4]$, where chaos is predominant. Each eigenstate is represented by a thin gray line. One sees that most eigenstates cluster slightly above $\Lambda_\alpha=1$, indicating that they are nearly as delocalized as random pure states. Values of $\Lambda_\alpha<1$ are possible, but disappear as $j$ increases~\cite{Pilatowsky2021NatCommun}. A portion of the eigenstates has values of $\Lambda_\alpha$ that are much higher than 1, some reaching $\Lambda_4=5$. As we show next, these high values of localization are caused by strong quantum scarring. It is worth noting, however, that the lines in Fig.~\ref{fig01} are not parallel and rather have multiple crossings. This tells us that there is indeed information to be gained by analyzing $\Lambda_\alpha$ for different values of $\alpha$.

The thick colored lines in Fig.~\ref{fig01} mark 8 eigenstates, labeled A-H, that we select for further analysis. These eigenstates constitute a representative sample of the different behaviors we see in Fig.~\ref{fig01}. States A and B have the highest value of $\Lambda_\alpha$ for $\alpha>2$, D has the highest value of $\Lambda_\alpha$ for $\alpha<1$, H has the lowest value of $\Lambda_\alpha$ for all $\alpha$, and $\Lambda_\alpha(G)=1$ for all $\alpha$. States C, E, and F were selected to have evenly spread values of $\Lambda_4$ between B and G. We focus on these 8 eigenstates henceforth, but we have indeed verified that the analysis we perform describes other eigenstates with similar values of $\Lambda_\alpha$, and hence the conclusions we reach are general.

\begin{figure*}[p]
\centering
\includegraphics[width=0.98\textwidth]{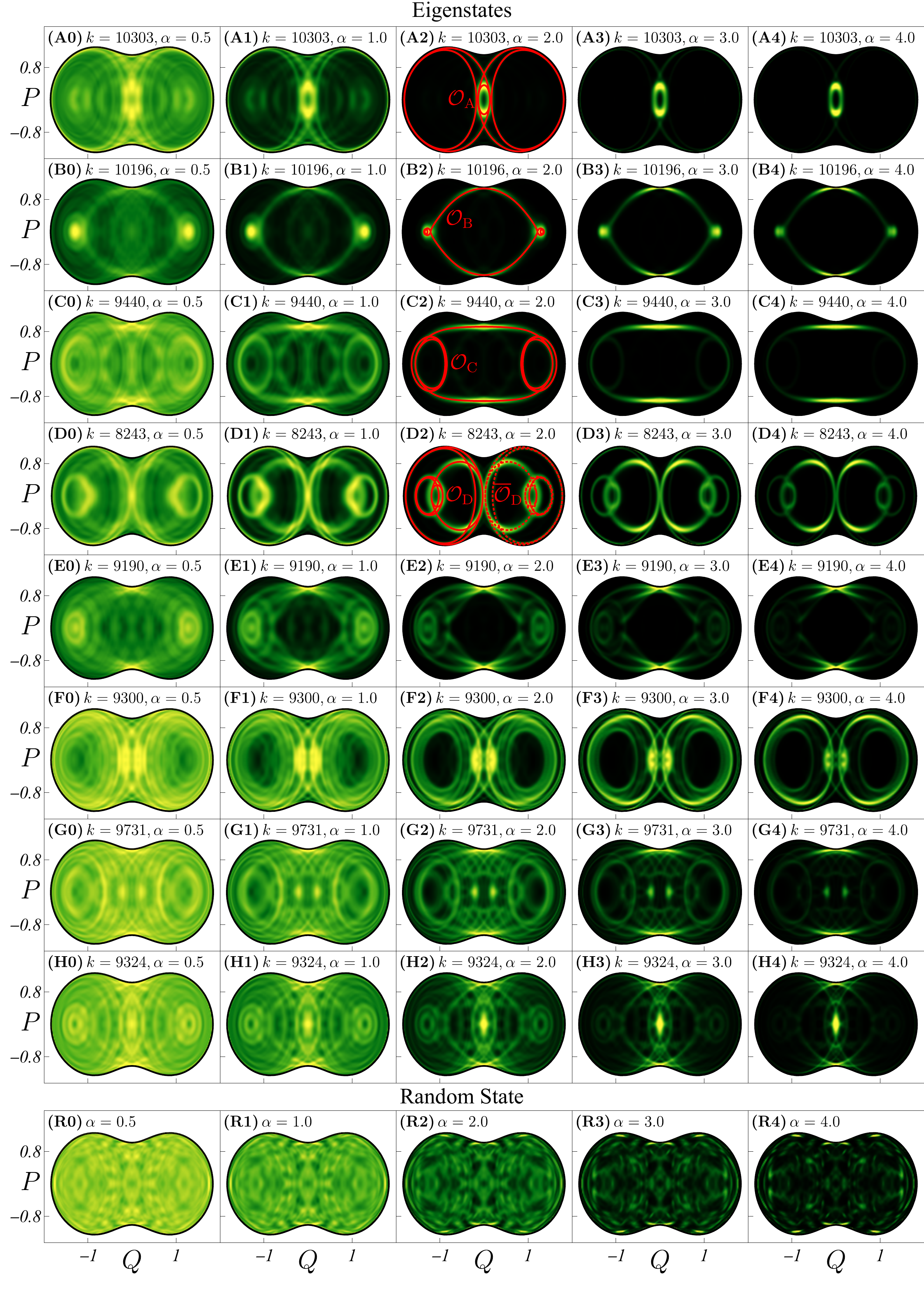}\vspace{-14px}
\caption{Projected moments of the Husimi functions, $\widetilde{\mathcal{Q}^\alpha_k}(Q,P)$, for the eigenstates labeled A-H in Fig.~\ref{fig01} and $\widetilde{\mathcal{Q}^{\alpha}_{\hat{\rho}_R} } (Q,P, \epsilon)$ for a pure random state (R) from a Gaussian orthogonal ensemble obtained by weighting the positive parity eigenstates inside an energy window of width $1.5$ centered at the energy $\epsilon=-0.5$. Each column corresponds to a different value of $\alpha\in\{0.5,1,2,3,4\}$. The solid red line in (A2)-(D2) represent the projections in $(Q,P)$ of the unstable periodic orbits $\mathcal{O}_{\text{A}}$, $\mathcal{O}_{\text{B}}$, $\mathcal{O}_{\text{C}}$, $\mathcal{O}_{\text{D}}$, and the dashed red line in (D2) is for the mirrored orbit $\overline{\mathcal{O}}_{\text{D}}$. The system size is $j=100$.}
\label{fig02}
\end{figure*}

Visualizing the moments of the four-dimensional Husimi function comes to our aid for explaining the different degrees of localization of the eigenstates shown in Fig.~\ref{fig01}. We consider the projection into the atomic $(Q,P)$ plane of the $\alpha$-moments of the Husimi function $\mathcal{Q}_k$ of eigenstates $\hat{\rho}_k$ intersected with the classical energy shell at $\epsilon_k$, so that we obtain two-dimensional pictures, that is,
\begin{equation}
\widetilde{\mathcal{Q}^{\alpha}_{k} } (Q,P)  =  \iint \dif q\dif p \,\delta(h_\text{cl}({\bm x})-\epsilon_k)\, \mathcal{Q}_{k}({\bm x})^{\alpha}.
\end{equation}
In Refs.~\cite{Pilatowsky2021NatCommun,Pilatowsky2021}, we studied $\widetilde{\mathcal{Q}^1_k}(Q,P)$ and verified that it allows for the visual identification of quantum scars. In Ref.~\cite{Pilatowsky2021NatCommun}, we concluded that all the eigenstates of the Dicke model are scarred, even though most of them are almost as delocalized as random states. This is corroborated again by the representative eigenstates shown in Fig.~\ref{fig02}.

In Fig.~\ref{fig02}, we compare $\widetilde{\mathcal{Q}^{\alpha}_{k} } (Q,P)$ for the 8 representative eigenstates A-H from Fig.~\ref{fig01} and also $\widetilde{\mathcal{Q}^{\alpha}_{\hat{\rho}_R} } (Q,P, \epsilon)
= \iint \dif q\dif p \,\delta(h_\text{cl}({\bm x})-\epsilon)\, \mathcal{Q}_{\hat{\rho}_R}({\bm x})^{\alpha}
$ for a random state $\hat{\rho}_R$ centered at energy $\epsilon$. This state is obtained by weighting the positive parity eigenstates inside a rectangular energy window  with real coefficients normally distributed as in the Gaussian orthogonal ensembles of random matrix theory. For each state, we show the distributions for five values of  $\alpha\in\{0.5,1,2,3,4\}$. 

In contrast to the random state [Fig.~\ref{fig02}~(R0)-(R4)], the projected moments of the Husimi functions of the eigenstates [Fig.~\ref{fig02}~(A0)-(H4)] always display structures that are related with underlying unstable periodic orbits, and therefore imply that the eigenstates are scarred~\cite{Pilatowsky2021NatCommun}. When examining the figure, one should keep in mind that values of $\alpha<1$ tend to homogenize the distribution $\widetilde{\mathcal{Q}^\alpha_k}(Q,P)$, up to the limiting case where $\widetilde{\mathcal{Q}^0_k}(Q,P)=1$ independently of $(Q,P)$. On the other hand, values of $\alpha>1$ tend to erase the small contributions from the Husimi function, so that regions where $\widetilde{\mathcal{Q}^\alpha_k}(Q,P)$ is large get enhanced. The high $\alpha$-moments of the Husimi function are therefore useful tools in the analysis of quantum scarring, as discussed next.

\section{Quantum \,Scarring \,and \,Unstable Periodic Orbits}
\label{sec:3}

Quantum scarring corresponds to the concentration of a quantum state around the phase-space region occupied by a periodic orbit 
\begin{equation}
    \mathcal{O}=
    \{  \bm x(t)  \, \mid \, 
    t  \in [0,T]  \} \subseteq \mathcal{M}(\epsilon)
\end{equation}
with a periodic initial condition $\bm{x}(T) = \bm x(0)$
that  is unstable, that is, has a positive Lyapunov exponent $\lambda$. These orbits are always present in chaotic systems and are deeply connected with the quantum spectrum of such systems~\cite{Gutzwiller1971,Gutzwiller1990book}.

\subsection{Identifying Unstable Periodic Orbits}

By tracking the peaks of the projected fourth-moment of the Husimi function displayed in Figs.~\ref{fig02} (A4), (B4), (C4), and (D4), we can uncover the unstable periodic orbits that significantly scar the states A-D. The procedure goes as follows. Those peaks give a set of possible initial conditions for periodic orbits on the variables $(Q_i,P_i)$. By varying the bosonic coordinate $p$ and selecting $q$ according to the fixed energy, we identify the whole set of coordinates  $\bm x_i\in \mathcal{M}(\epsilon_k)$ where the Husimi function attains a maximum, thus yielding a set of possible initial conditions for a periodic orbit on all variables $\bm x_i$. We evolve these initial conditions and select those that roughly follow the green lines in Figs.~\ref{fig02} (A4), (B4), (C4), and (D4). The evolution is halted when the evolved point $\bm x_i(T_i)$ approximately returns to the initial condition $\bm x_i(0)$. This gives an approximate period $T_i$ and an initial condition $\bm x_i$, which can be converged to a truly  periodic condition by means of an algorithm known as the monodromy method~\cite{Baranger1988,DeAguiar1988,Simonovi1999, Pilatowsky2021}\footnote{Code available at  \href{https://github.com/saulpila/DickeModel.jl}{github.com/saulpila/DickeModel.jl}.}. This procedure leads to the five unstable periodic orbits labeled $\mathcal{O}_{\text{A}}$, $\mathcal{O}_{\text{B}}$, $\mathcal{O}_{\text{C}}$, $\mathcal{O}_{\text{D}}$, and $\overline{\mathcal{O}}_{\text{D}}$ in Figs.~\ref{fig02} (A2), (B2), (C2), and  (D2). The mirrored orbit $\overline{\mathcal{O}}_{\text{D}}$ is obtained by changing the signs of the coordinates $Q$ and $q$ of $\mathcal{O}_{\text{D}}$. We only mirror $\mathcal{O}_{\text{D}}$, because the other orbits are symmetric under the parity  transformation $(Q,q)\mapsto (-Q,-q)$. We stress that these unstable periodic orbits are different from those that we found in Ref.~\cite{Pilatowsky2021}, whose origin could be traced down to the fundamental excitations around the ground state configuration. Further studies of these new orbits may eventually reveal the properties and origin of these new families of unstable periodic orbits for the Dicke model.

The Lyapunov times $1/\lambda$ of the unstable periodic orbits identified here are smaller but of the same order of  their respective periods $T$. Table~\ref{tab:01} gives the values of the Lyapunov exponents, $\lambda$, and the periods divided by the Lyapunov times, $T\lambda$, for each of the five unstable periodic orbits that we found. The competition between the period of the orbit and the Lyapunov time is important for scarring. As originally discussed in Ref. \cite{Heller1984}, a non-stationary state will be affected by a neighboring periodic orbit if the Lyapunov time of the orbit is larger than its period. If instead the Lyapunov time is shorter than the period, the periodic orbit does not have time to develop and scar the non-stationary state. Here, we observe that even if this condition is slightly broken, $T\lambda\gtrsim 1$,  the periodic orbits still cause a significant localization of the eigenstates.  

\begin{table}
\centering
\begin{tabular}{|c|c|c|c|c|}
\hline\hline 
  \quad & $\mathcal{O}_{\text{A}}$  & $\mathcal{O}_{\text{B}}$ & $\mathcal{O}_{\text{C}}$  & $\mathcal{O}_{\text{D}},\overline{\mathcal{O}}_{\text{D}}$ \\[1ex]
   \hline\hline
    & & & &\\[-1ex]
  $\lambda$ & 0.143 & 0.191 & 0.253 &  0.153 \\[1ex]
  \hline
    & & & &\\[-1ex]
  $T\lambda$ & 1.86 & 2.97 & 3.43 &  2.21 \\[1ex]
  \hline
    & & & &\\[-1ex]
  $\mathcal{P}_k$ & 33.2 & 31.1 & 28.3 &  22.5 \\[1ex]
 \hline\hline
\end{tabular}
\caption{\label{tab:01}
Lyapunov exponent $\lambda$, its product with the period $T$, and the scarring measure $\mathcal{P}_{k}(\mathcal{O})$ \eqref{eqn:scarmeas} for the orbits A-D.}
\end{table}

To verify that the states A-D are indeed scarred by the unstable periodic orbits $\mathcal{O}_{\text{A}}$-$\mathcal{O}_{\text{D}}$, we employ the following scarring measure~\cite{Pilatowsky2021}:
\begin{equation}
\label{eqn:scarmeas}
\mathcal{P}_k(\mathcal{O})=\frac{\tr(\hat{\rho}_k \hat{\rho}_{\mathcal{O}})}{\tr( \hat{\rho}_\epsilon \hat{\rho}_{\mathcal{O}})},
\end{equation}
where 
\begin{equation}
\label{eq:tubestate}
    \hat{\rho}_{\mathcal{O}}=\frac{1}{T}\int_0^{T}\dif t\, \dyad{\bm x(t)}
\end{equation} 
is a tubular state composed of all coherent states $\ket*{\bm x(t)}$ whose centers lie in the periodic orbit $\mathcal{O}=\{\bm x(t) \mid t\in[0,T)\}$ and $ \hat{\rho}_\epsilon=\energyshellaveragelong{\dyad{\bm x}}{\bm x}{\epsilon}$ is a mixed state completely delocalized in the classical energy shell at $\epsilon=h_\text{cl}(\mathcal{O})$, that is, a mixture of all coherent states whose centers lie within this classical energy shell (see details in Ref.~\cite{Pilatowsky2021}). A value $\mathcal{P}_k(\mathcal{O})=n$ indicates that the state $\hat{\rho}_k=\dyad{E_k}$ is $n$ times more likely to be found near the orbit $\mathcal{O}$ than a completely delocalized state, for which $\mathcal{P}_k(\mathcal{O})=1$. For eigenstates not scarred by the periodic orbit $\mathcal{O}$ and also random states,  $\mathcal{P}_k(\mathcal{O})\leq 1$ \cite{Pilatowsky2021}. The results of the scarring measure obtained with Eq.~\eqref{eqn:scarmeas} for the states A-D are shown in Table~\ref{tab:01} and confirm that these states are significantly scarred by the unstable periodic orbits $\mathcal{O}_{\text{A}}$-$\mathcal{O}_{\text{D}}$.

One observes secondary structures in Figs.~\ref{fig02}~(A0), (B0), (C0), and (D0) that are not directly generated by the unstable periodic orbits delineated in Figs.~\ref{fig02}~(A2), (B2), (C2), and (D2). These structures may be related to the self-interferences of the unstable periodic orbits~\cite{Berry1989} or to other secondary scars associated with other unstable periodic orbits that we have not identified. We leave this question for future investigations.

\begin{figure*}[ht]
\centering
\includegraphics[width=1\textwidth]{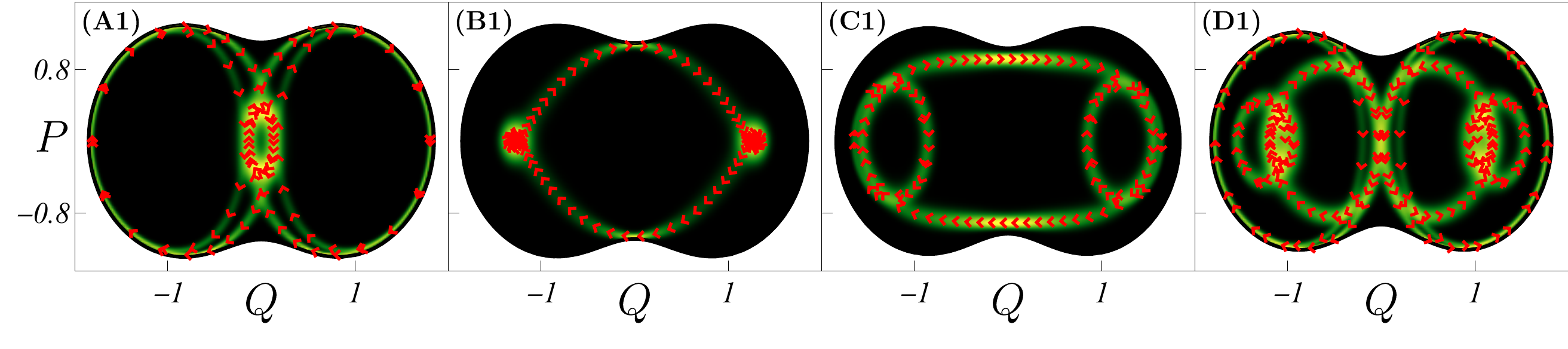}
\includegraphics[width=1\textwidth]{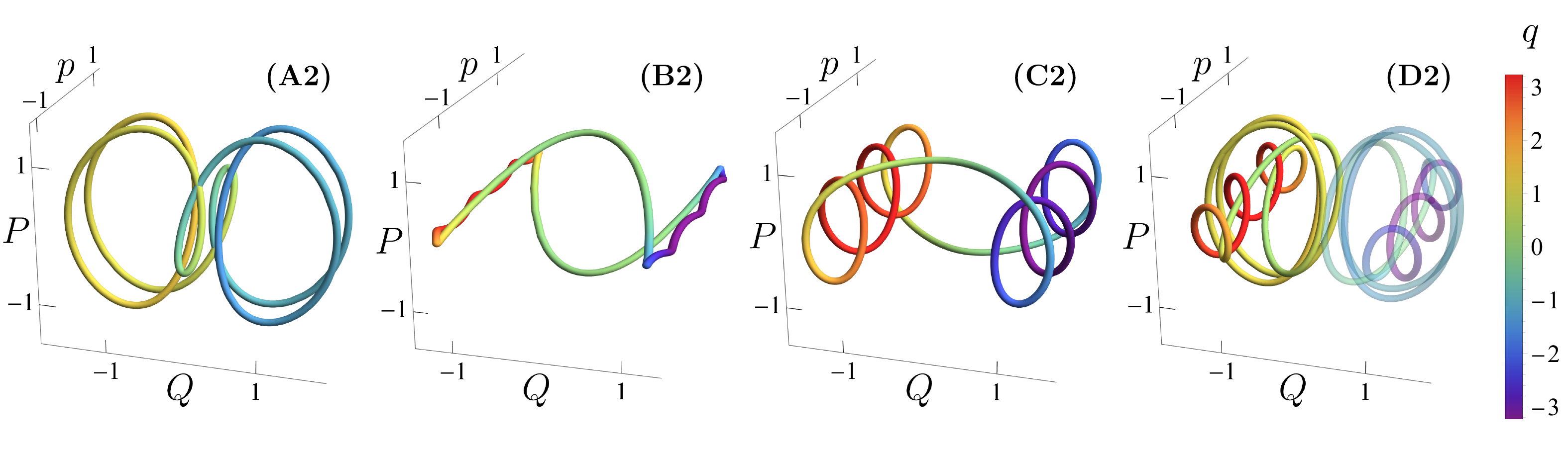}
\caption{Top panels: in green, the Husimi projection $\widetilde{\mathcal{Q}}_\mathcal{O}$ for the orbits $\mathcal{O}_{\text{A}}$ (A1), $\mathcal{O}_{\text{B}}$ (B1), $\mathcal{O}_{\text{C}}$ (C1), and added  projections $(\widetilde{\mathcal{Q}}_{\mathcal{O}_{\text{D}}}+\widetilde{\mathcal{Q}}_{\overline{\mathcal{O}}_{\text{D}}})/2$ for orbit $\mathcal{O}_{\text{D}}$ and the mirrored orbit  $\overline{\mathcal{O}}_{\text{D}}$ (D1). The red arrows are placed at constant time intervals along the unstable periodic orbits  $\mathcal{O}_{\text{A}}$ (A1), $\mathcal{O}_{\text{B}}$ (B1), $\mathcal{O}_{\text{C}}$ (C1), and $\mathcal{O}_{\text{D}},\overline{\mathcal{O}}_{\text{D}}$ (D1). Bottom panels: Three-dimensional plots of the periodic orbits $\mathcal{O}_{\text{A}}$ (A2), $\mathcal{O}_{\text{B}}$ (B2), $\mathcal{O}_{\text{C}}$ (C2), and $\mathcal{O}_{\text{D}}$ (opaque) $\overline{\mathcal{O}}_{\text{D}}$ (translucent) (D2). In these panels, the variables $Q$, $P$, and $p$ correspond to each of the three axis, as marked, and the color represents the fourth variable $q$. The system size is $j=100$.}
\label{fig03}
\end{figure*}

\subsection{Highly localized eigenstates and  scarring}

The dynamical properties of the unstable periodic orbits that scar states A-D allow us to explain the structures of the distributions in Figs.~\ref{fig02} (A0)-(D4). To do this, we consider the Husimi function of the tubular state $\hat{\rho}_\mathcal{O}$,
\begin{equation}
    \mathcal{Q}_\mathcal{O}(\bm x)=\expval{\hat{\rho}_\mathcal{O}}{\bm x}=\frac{1}{T}\int_0^{T}\dif t\, \abs{\braket{\bm x}{\bm y(t)}}^2,
\end{equation}
where $\bm y(t)\in \mathcal{O}$, and compute the projection of this function into the $(Q,P)$ plane,
\begin{equation}
    \widetilde{\mathcal{Q}}_\mathcal{O}(Q,P)=\iint \dif q\dif p \,\delta(h_\text{cl}(\bm x)-\epsilon)\, \mathcal{Q}_\mathcal{O}(\bm x), 
\end{equation}
at energy $\epsilon=h_\text{cl}(\mathcal{O})$. In Figs.~\ref{fig03}~(A1), (B1), and (C1) we plot in green  the projections $\widetilde{\mathcal{Q}}_\mathcal{O}$ for the orbits $\mathcal{O}_{\text{A}}$, $\mathcal{O}_{\text{B}}$, and $\mathcal{O}_{\text{C}}$, respectively, and in Fig.~\ref{fig03}~(D1) we plot the added projections $(\widetilde{\mathcal{Q}}_{\mathcal{O}_{\text{D}}}+\widetilde{\mathcal{Q}}_{\overline{\mathcal{O}}_{\text{D}}})/2$ for orbit $\mathcal{O}_{\text{D}}$ and the mirrored orbit  $\overline{\mathcal{O}}_{\text{D}}$.  One sees that the projections  $\widetilde{\mathcal{Q}}_\mathcal{O}(Q,P)$ of the Husimi function of the tubular state in Figs.~\ref{fig03}~(A1), (B1), (C1), and (D1) are similar to the projections  $\widetilde{\mathcal{Q}^{\alpha}_{k} } (Q,P)$ of the Husimi function of the eigenstates A-D shown in Figs.~\ref{fig02}~(A2)-(D2), Figs.~\ref{fig02}~(A3)-(D3), Figs.~\ref{fig02}~(A4)-(D4). 

In Figs.~\ref{fig03}~(A2)-(D2), we show three-dimensional plots of the four-dimensional periodic orbits $\mathcal{O}_{\text{A}}$, $\mathcal{O}_{\text{B}}$, $\mathcal{O}_{\text{C}}$, and both $\mathcal{O}_{\text{D}}$ and $\overline{\mathcal{O}}_{\text{D}}$. The  plots display the variables $Q$, $P$, and $p$, and the color represents the fourth variable $q$. The shapes seen in Figs.~\ref{fig03}~(A1)-(D1) are precisely the orbits in Figs.~\ref{fig03}~(A2)-(D2) projected into the $(Q,P)$-plane.

Figure~\ref{fig03} contains only semiclassical information, which helps us understand specific features and the degree of localization of the eigenstates A-D. For example, when the classical dynamics is fast, the scars are less intense. In  Figs.~\ref{fig03}~(A1)-(D1), we place red arrows at constant time intervals along the corresponding periodic orbits. The closer those arrows are, the slower the classical dynamics is. One sees that the brighter green regions correspond to the regions with a high density of red arrows. These are regions of longer permanence of the classical orbit, which produce deeper imprinted scars \cite{Pilatowsky2021NatCommun, Pilatowsky2021, Shane2020}. 

The dynamical properties of the orbits also explain why  $\Lambda_{\alpha}$ increases with $\alpha$ for some eigenstates. The faster the dynamics is in a given region of the phase space, the less probable it is to find a point following the orbit there. This gets reflected in the high moments ($\alpha>1$) of the Husimi function, where small contributions tend to be erased, increasing the value $\Lambda_{\alpha}$. We can see this effect by comparing Figs. \ref{fig02} and \ref{fig03}. As $\alpha$ grows, some colored regions in the leftmost columns of Fig. \ref{fig02} are not visible in the rightmost columns. These are regions of low probability, which have faster dynamics (more separated arrows in Fig. \ref{fig03}).

The general features discussed in the two previous paragraphs are described below in more detail for the eigenstates A-D.

\begin{itemize}
     \item The unstable periodic orbit $\mathcal{O}_{\text{A}}$ [Fig.~\ref{fig03}~(A1)], which heavily scars state A, has slower dynamics near the small oval region in the center of the orbit, where most red arrows are concentrated. Comparing $\mathcal{O}_{\text{A}}$ with $\mathcal{O}_{\text{B}}$, one sees that $\mathcal{O}_{\text{A}}$ is  larger in phase space, so state A is less localized than state B and $\Lambda_\alpha(A)<\Lambda_\alpha(B)$ for $\alpha<3$, as shown in Fig.~\ref{fig01}. Nevertheless, $\alpha=4$ is sufficiently large to erase the less bright regions of $\mathcal{O}_{\text{A}}$, where the classical dynamics is fast, leaving only the contributions from the central oval region and thus leading to a higher value of localization and to $\Lambda_\alpha(A) > \Lambda_\alpha(B)$ for $\alpha>3$.
    \item The unstable periodic orbit $\mathcal{O}_{\text{B}}$ covers only a small portion of the classical energy shell and has a relatively constant speed in the phase space, as seen by the constant density of red arrows in Fig.~\ref{fig03} (B1), so the state B maintains a high value of localization in Fig.~\ref{fig01} for all values of $\alpha\in[0,4]$.
    \item The unstable periodic orbit $\mathcal{O}_{\text{C}}$ presents two bright regions in Fig.~\ref{fig03}~(C1), at $Q=0$ and $P\approx \pm0.8$, where the dynamics is slow. As seen in Fig.~\ref{fig01}, state C has a lower value of $\Lambda_\alpha$ than state B, because $\mathcal{O}_{\text{C}}$ is larger in phase-space than $\mathcal{O}_{\text{B}}$. However, the slope of $\Lambda_\alpha$ at $\alpha=4$ is larger for state C than for state B, because the less bright loops of $\mathcal{O}_{\text{C}}$, where the classical dynamics is fast, disappear for $\alpha=4$ [Fig.~\ref{fig02}~(C4)], leaving only the two bright spots at $Q=0$.
    \item The behavior of $\Lambda_\alpha$ for state D in Fig.~\ref{fig01} is peculiar in that it initially grows for $\alpha<2$, but flattens after $\alpha>2$. The localization measure increases as $\alpha$ grows to 2, because the contribution of the fast outer loops strongly diminish. But then, for $\alpha>2$, since the dynamics in the loops of intermediate sizes has a relatively constant speed, $\Lambda_\alpha$ flattens. In fact, Fig.~\ref{fig01} suggests that the flattening for large $\alpha$ holds also for state B, while $\Lambda_\alpha$ for states A and C keeps growing at least up to $\alpha=4$.
\end{itemize}

We close this subsection by discussing state E, which in terms of localization is midway between the highly localized eigenstates A-D and the maximally delocalized eigenstates F-H described in the next subsection. The projected Husimi moment distributions depicted in Figs.~\ref{fig02}~(E0)-(E4) show localization around some regions that must be associated with a classical unstable periodic orbit with slow dynamics at $Q\approx 0$ and $P\approx \pm 0.9$. However, the numerical method used to identify the orbits that scar the eigenstates A-D fails for state E. This suggests  that the ratio between the period and the Lyapunov time, $T\lambda$, of the unstable periodic orbit that scars this eigenstate should be much larger than one.  

\subsection{Maximally delocalized eigenstates}

As seen in Fig.~\ref{fig01}, the eigenstates F, G and H have values of $\Lambda_\alpha\approx 1$, similar to
those of random states. However, contrary to Figs.~\ref{fig02}~(R0)-(R4), the panels (F0)-(H4) show that these eigenstates still display orbit-like patterns, that is, they exhibit closed loops. These patterns must belong to a set of periodic orbits that scar the eigenstates~\cite{Pilatowsky2021NatCommun}, but we have not yet been able to identify these orbits with the  numerical method that we have implemented. They must have periods that are much larger than their corresponding Lyapunov times, which makes them more difficult to find. Since these states are scarred by unstable periodic orbits of larger periods, they are more delocalized than the states A-D, which explains why $\Lambda_\alpha(F),\Lambda_\alpha(G),\Lambda_\alpha(H)<\Lambda_\alpha(A), \Lambda_\alpha(B), \Lambda_\alpha(C), \Lambda_\alpha(D)$. 

The results in this work support our claims in Ref.~\cite{Pilatowsky2021NatCommun} that all eigenstates are scarred by unstable periodic orbits, yet some of them are as delocalized in phase space as random states. Identifying the unstable periodic orbits that scar the states E-H would bring further confirmation to this conjecture.

\section{Summary and Conclusions}
\label{sec:Conclusions}

Generalized inverse participation ratios and R\'enyi entropies of order $\alpha$ quantify the degree of delocalization of quantum states in Hilbert space. Our interest in this work was instead in the structure of quantum states in phase space. For this analysis, we employed a measure of localization called R\'enyi occupation of order $\alpha$, which is defined over classical energy shells and is based on the $\alpha$-moments of the Husimi function.
We showed that the R\'enyi occupations of random states have a simple analytical form that is a function of $\alpha$ only. We used this result to define a localization measure given by the ratio of the energy-shell occupation of random states to that of the state under investigation. This measure was then used to analyze the eigenstates of the Dicke model.

As $\alpha$ increases beyond 1, the smaller contributions from the Husimi function are erased and only the larger peaks survive. By examining these peaks in the fourth-moment of the Husimi function of highly localized eigenstates, we were able to visually identify the classical unstable periodic orbits that scar those states.

Our study of the classical dynamics of the identified orbits assisted our understanding of the structure and the degree of localization of the eigenstates. In regions of the phase space where the classical dynamics is slow, the scars are deeper imprinted, so the values of the $\alpha$-moments of the Husimi functions are larger and persist as $\alpha $ increases.

In conclusion, we have shown that quantum states supply valuable information about the non-linear dynamics of the classical limit of the model. Highly localized states are scarred by orbits of relatively short period and the peaks in the high-moments ($\alpha>1$) of their Husimi distributions provide a powerful tool to identify classical unstable periodic orbits. This is a major accomplishment given the difficulty in identifying the unstable periodic orbits underlying scarred states. 
Our analysis shows that we can learn about the classical dynamics by studying the quantum realm.

The localization measure $\Lambda_{\alpha}$ introduced in this work applies to other models with unbounded phase space, such as quantum billiards, and our technique to find classical unstable periodic orbits could be extended to those systems as well.

\section*{Acknowledgments}
We acknowledge the support of the Computation Center - ICN, in particular to Enrique Palacios, Luciano D\'iaz, and Eduardo Murrieta. SP-C, DV, and JGH acknowledge financial support from the DGAPA- UNAM project IN104020, and SL-H from the Mexican CONACyT project CB2015-01/255702. LFS was supported by the NSF Grant No. DMR-1936006.

\appendix

\section{Appendix: Semiclassical Density of States}
\label{app:SDoS}
The semiclassical approximation of the density of states of a quantum system is obtained by taking the first term of the well-known Gutzwiller trace formula~\cite{,Gutzwiller1971,Gutzwiller1990book}. This semiclassical density of states is the ratio of the three-dimensional volume of the classical energy shell to the four-dimensional volume of the Planck cell,
\begin{equation}
\nu(\epsilon) =  \frac{1}{(2\pi\hbar_{\text{eff}})^2}\int_{{\mathcal{M}}} \dif{\boldsymbol{x}}\, \delta(h_{\text{cl}}({\boldsymbol{x}})-\epsilon).
\end{equation}
The explicit expression of $\nu(\epsilon)$ for the Dicke model was derived in Ref.~\cite{Bastarrachea2014a}. Here, this expression is slightly modified to be consistent with the notation used in this article, and it reads
\begin{align}
\nu(\epsilon)  = \frac{2j^2}{\omega}\left\{\begin{array}{ll}
   \frac{1}{\pi}\int_{y_{-}}^{y_{+}} dy f(y,\epsilon)  & \hspace{-30px}\text{if } \epsilon_0 \leq \epsilon \leq -\omega_0, \\
   \frac{1+\epsilon/\omega_0}{2}+\frac{1}{\pi}\int_{\epsilon/\omega_0}^{y_{+}} dy f(y,\epsilon)&\hspace{-5px}\text{if }  |\epsilon| <\omega_0, \\
   1  &\text{if } \epsilon \geq \omega_0 ,
\end{array}\right.    
\end{align}
where
\begin{equation}
f(y,\epsilon)=\arccos\left(\sqrt{\frac{2\gamma_c^2(y-\epsilon/\omega_0)}{\gamma^2(1-y^2)}}\right),
\end{equation}
\begin{equation}
y_{\pm}=-\frac{\gamma_c}{\gamma}\left(\frac{\gamma_c}{\gamma} \mp \sqrt{\frac{2(\epsilon-\epsilon_0)}{\omega_0}}\right), 
\end{equation}
and
\begin{equation}
\epsilon_0=-\frac{1}{2}\left(\frac{\gamma_c^2}{\gamma^2}+\frac{\gamma^2}{\gamma_c^2}\right).    
\end{equation}

\bibliography{bibliography}

\end{document}